\documentclass[3p,times,preprint,authoryear]{elsarticle}

\usepackage{lineno,hyperref}
\modulolinenumbers[5]

\usepackage{aas_macros}

\usepackage{graphics}
\usepackage{graphicx}
\usepackage{epsfig}

\usepackage{amssymb}
\usepackage{rotating}
\usepackage{lscape}

\journal{Advances in Space Research}

\begin{document}

\begin{frontmatter}



\title{Hard X-ray Cataclysmic Variables}


\author[label1]{D. de Martino\corref{cor}}
\cortext[cor]{Corresponding author}
\ead{domitilla.demartino@inaf.it}

\author[label2,label1,label3]{F. Bernardini}
\ead{federico.bernardini@inaf.it}

\author[label4,label5]{K. Mukai}
\ead{Koji.Mukai@nasa.gov}

\author[label6,label7]{M. Falanga}
\ead{mfalanga@issibern.ch}

\author[label8,label9]{N. Masetti}
\ead{nicola.masetti@inaf.it}

\address[label1]{Istituto Nazionale di Astrofisica, Osservatorio Astronomico di Capodimonte, 
Salita Moiariello 16, I-80131, Napoli, Italy}
\address[label2]{Istituto Nazionale di Astrofisica, Osservatorio Astronomico di Roma,
Via  Frascati 33, I-00040,  Monte Porzio Catone (RM), Italy}
\address[label3]{New York University Abu Dhabi, Saadiyat Island, Abu Dhabi 129188, United 
Arab Emirates}
\address[label4]{CRESST and X-ray Astrophysics Laboratory, NASA Goddard Space Flight Center,
Greenbelt, MD 20771, USA}
\address[label5]{Department of Physics, University of Maryland, Baltimore County, 1000 
Hilltop Circle, Baltimore, MD 21250, USA}
\address[label6]{International Space Science Institute (ISSI), Hallerstrasse 6, 
CH-3012 Bern, Switzerland}
\address[label7]{International Space Science Institute Beijing, No.1 Nanertiao, 
Zhongguancun, Haidian District, 100190 Beijing, China}
\address[label8]{Istituto Nazionale di Astrofisica, Osservatorio di Astrofisica e 
Scienza dello Spazio di Bologna, Via Gobetti 93/3, I-40129, Bologna, Italy}
\address[label9]{Departamento de Ciencias F\'isicas, Universidad Andr\'es Bello,
Fern\'andez Concha 700, Las Condes, Santiago, Chile }

\begin{abstract}
Among hard X-ray Galactic sources detected in the  {\it Swift} and  {\it INTEGRAL} 
surveys,
those discovered as accreting white dwarf binaries have suprisingly boosted in number
in the recent years. The majority are identified as magnetic Cataclysmic Variables
of the Intermediate Polar type, suggesting this subclass as an important
constituent of the Galactic population of X-ray sources.  We here review and discuss 
the  X-ray emission properties of newly discovered sources in the framework of an
identification programme with the {\it XMM-Newton} satellite that 
increased  the sample of this subclass by a factor of two. 

 \end{abstract}


\end{frontmatter}

\parindent=0.5 cm

\section{Introduction}

Cataclysmic Variables (CVs) are low-mass close binary systems composed by a white dwarf (WD) primary 
accreting material lost from a Roche-lobe filling, late-type companion star.  
According to the variety of observed phenomenology they are grouped in 
several subclasses \citep[see review][]{warner}. Two main categories can
be identified: the magnetic systems (mCVs) harbouring WD primaries with
strong  magnetic fields ($\rm B_{WD} \geq 10^6\,G$ ) and non-magnetic CVs.
The magnetic systems are also subdivided in two groups, the polars that
contain  highly ($\rm B_{WD} \sim 14-230\times 10^6\,G$) 
magnetised WDs, and the intermediate polars (IPs)
believed to
harbour low magnetic field primaries ($\rm B_{WD} \lesssim 10^7\,G$) 
 \citep[see reviews by][]{cropper90,ferrario15,mukai17}. In both subclasses
the WD magnetic field plays a key role in shaping the accretion flow.
In the polars the B-field is strong enough to lock, or quasi-lock, the WD 
rotation at the  binary orbit, and to channel matter from the donor star 
onto the WD polar caps through an accretion stream.  Thus, polars show  
strong periodic orbital 
variability at all  wavelengths \citep[e.g.][]{schwope98}. 
In IPs, instead, the weaker field is not able to synchronise the WD rotation
at the binary orbit allowing fast spin rates ($\rm P_{spin=\omega} << P_{orb=\Omega}\sim 
hrs$) and the formation of an accretion disc truncated at the magnetospheric boundary
where the magnetic pressure balances the ram pressure. Depending on system
parameters, such as orbital period, magnetic field strength and mass accretion rate,
IPs may accrete via a disc (disc-fed systems), without a disc (disc-less) or in a  
hybrid mode in the form of disc overflow, which can be diagnosed through
the presence of spin, orbital and sidebands periodicities at different wavelengths
\citep{hellier95,norton96,norton97}.

\noindent The accretion flow close to the WD surface
is channeled along the magnetic field lines, reaching supersonic velocities 
and producing a stand-off shock above the WD surface \citep{aizu73}. 
The post-shock region (PSR) is hot  ($kT\sim10-50$ keV) and cools via thermal 
Bremsstrahlung (hard X-rays) and cyclotron radiation, emerging in the 
optical/nIR band.  Both emissions are partially thermalized by the WD surface and 
re-emitted in the soft X-rays and/or EUV/UV domains.
 The relative proportion
of the two cooling mechanisms strongly depends on the B-field strength and local
mass accretion rate. Cyclotron radiation dominates in the high field polars
and is efficient in suppressing high PSR temperatures  
\citep{woelk96,fischer01}.
An optically thick soft X-ray (kT$_{\rm bb} \sim$ 30-50\,eV) emission due to
reprocessing  \citep{van_teeseling94} or to heating due to blobby 
accretion \citep{woelk92}   was also found to
be strong in polars, explaining why they where found numerous 
in previous soft X-ray 
surveys (e.g.{\it ROSAT}), largely outnumbering IPs \citep{beuermann99,schwope02}. 
The PSR  has been diagnosed in details through circular and linear polarimetry
and spectro-po\-la\-ri\-me\-try at optical/nIR wavelengths,
in the polars only, revealing complex field topology with differences between the
primary and secondary poles \citep[e.g.][]{potter04,beuermann07,ferrario15}. In some 
cases differences in the magnetic field strengths have also been found
between the PSR region and the WD photosphere, the latter inferred
from Zeeman splitting of hydrogen lines  \citep[e.g.][]{schwope95,ferrario15}. 
On the contrary, most IPs do not show optical/nIR
polarization, with only 11 systems found to be polarised at a few
percent \citep[see][]{ferrario15,potter18}.
Their magnetic field strenghts are consequently difficult to measure
and are loosely estimated in the range 
$\rm \sim 5-30\times 10^{6}$\,G. These polarised IPs are all found at 
long orbital periods,  above the 2-3\,h orbital period gap, and could 
represent the long-sought progenitors of low-field polars that will 
evolve into synchronism.

\noindent The complex geometry and emission properties of mCVs make these low-mass
X-ray binaries ideal laboratories to study in details accretion processes in 
moderate magnetic field environments, 
but also help in understanding the role of magnetic fields in 
close-binary evolution. Indeed,  the incidence  of magnetism among CVs is  $\sim25\%$,
compared to $\sim 6-10\%$ of isolated magnetic WDs \citep{ferrario15}. 
This would either imply CV formation
is favoured by magnetism or CV production enhances magnetism \citep{tout08}.
In addition, mCVs are the brightest X-ray emitting CVs, with X-ray 
luminosities ranging from a few $\rm \sim 10^{30}\, erg\,s^{-1}$ to 
$\rm \sim10^{34}\,erg\,s^{-1}$, and may play a crucial
role in understanding Galactic X-ray binary populations.

\noindent We here review the recent progresses on mCVs obtained in the
framework of an identification programme with the {\it XMM-Newton} mission, 
aiming at classifying new mCV candidates discovered in the hard X-ray surveys
conducted by the   {\it Swift} and {\it INTEGRAL} satellites.
In Sect.\,2 we report on the outcomes from these surveys. In Sect.\,3 
we summarize  the new identifications, their temporal and spectral properties and in Sect.\,4
the role of fundamental parameters.  
In Sect.\,5 we conclude on the perspectives with future missions foreseen in the 2020-2030 timeframe.

\section{Cataclysmic Variables in hard X-ray surveys}\label{sec:surveys}

\begin{figure}
\centerline{
\includegraphics[width=1.1\columnwidth]{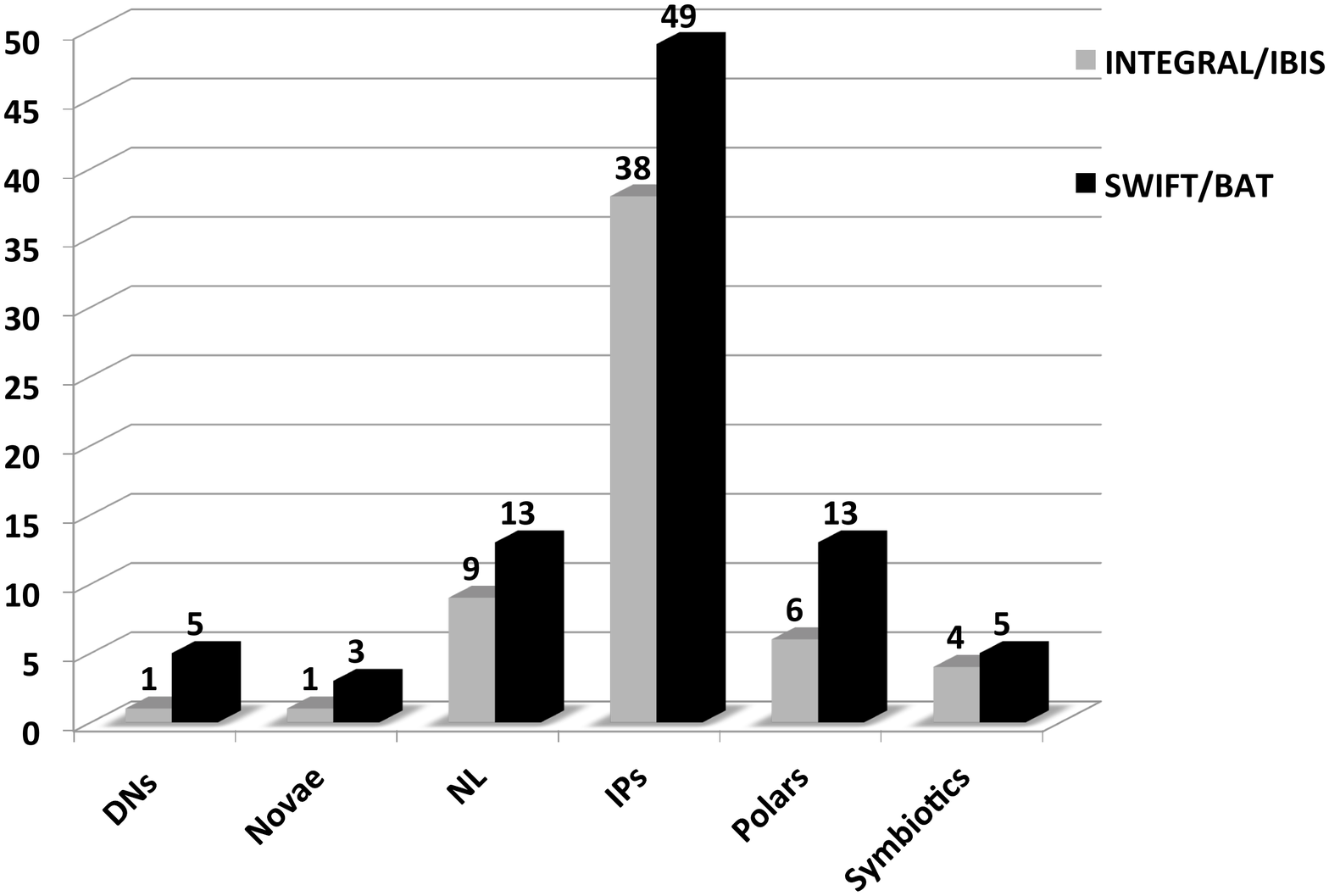}}
\caption{ The distribution of CV types detected by {\it INTEGRAL}/IBIS-ISGRI and 
{\it Swift}/BAT using the latest catalogue releases (misidentifications were
corrected). 
 \label{fig:fig1}}
\end{figure}

Our  view of the hard X-ray sky  dramatically changed thanks
to the deep  {\it INTEGRAL}/IBIS-ISGRI \citep{bird16} and {\it Swift}/BAT \citep{oh18}
surveys with more than 1600 sources detected above 20keV.
These surveys have shown that our knowledge of the X-ray
binary populations in the Galaxy was poor, surprisingly detecting 
a large number of accreting WD binaries,  amounting to $\sim 25\%$ of 
the Galactic sources. Both the {\it INTEGRAL}/IBIS-ISGRI 
Galactic plane survey \citep{barlow06,krivonos12,bird16}
and the {\it Swift/BAT} survey \citep{oh18}, mainly covering 
high Galactic latitudes, 
reveal a high incidence of mCVs with $\sim 70\%$  of them belonging to the 
IP group. In Fig.\,\ref{fig:fig1} we show the
accreting WD binaries detected from both surveys, comprising
of  a handful of  previously known non-magnetic Dwarf Novae  (DNs), old 
Novae and Symbiotics, many
Nova-like systems (NLs) (most are still disputed to be magnetic), and 
the magnetic IPs and polars. This finding indicates that the subclass of 
IP-type mCVs is not as small as previously believed, suggesting them as 
potential  important contributors to the Galactic X-ray source population. 
Indeed, 
the deep surveys of the Galactic centre conducted with {\it Chandra} 
\citep{muno04}, {\it XMM-Newton} \citep{heard13} and recently with 
{\it NuSTAR}, that overcomes the selection bias towards high temperatures of the
{\it INTEGRAL}/IBIS-ISGRI and {\it Swift}/BAT surveys,  
\citep{perez15,hailey16,hong16} all suggest the dominance
of mCVs of the IP-type above $\sim 10^{31}\rm erg\,s^{-1}$.
Whether these mCVs  also dominate the Galactic ridge
emission (GRXE) is still  disputed based on {\it RXTE} 
\citep{revnivtsev06}, {\it Chandra} \citep{revnivtsev09}, 
{\it XMM-Newton}  \citep{warwick14} and
{\it Suzaku} \citep{xu16,nobukawa16} observations.


\begin{figure}
\centerline{
\includegraphics[width=0.75\columnwidth,angle=-90]{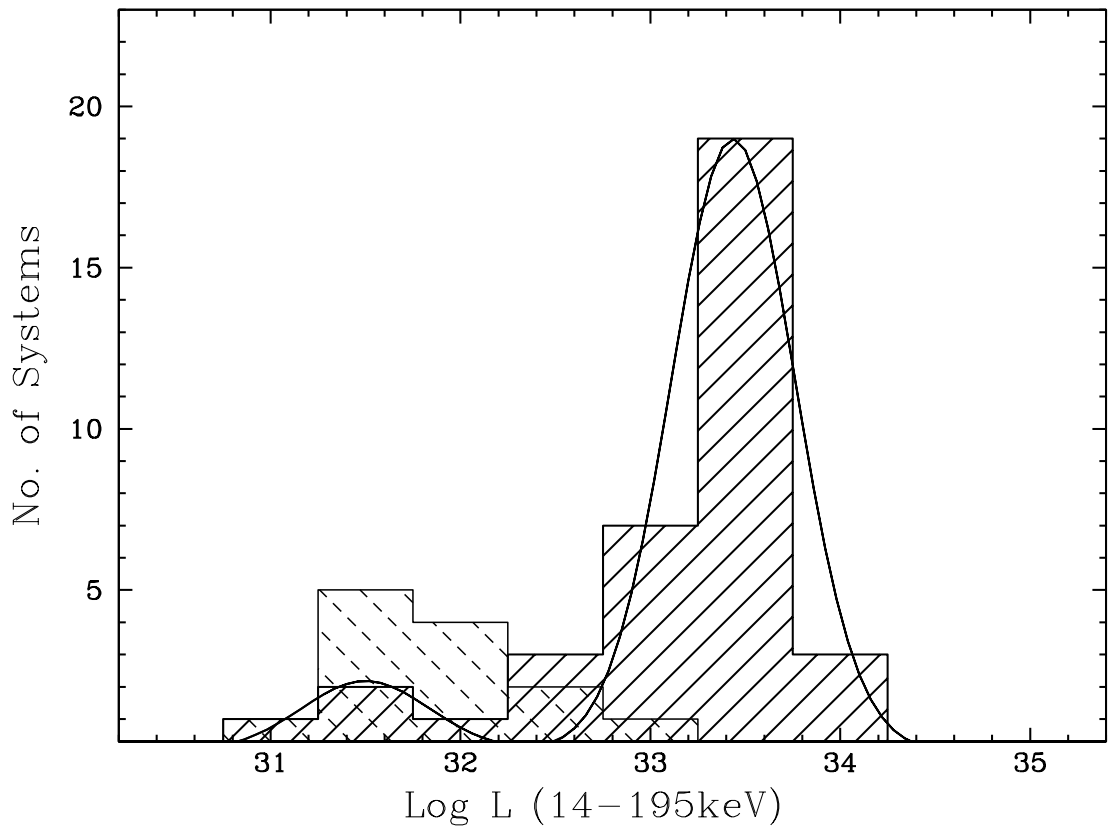}}
\caption{ The 14-195\,keV luminosity distribution of confirmed IPs (solid) and polars 
(dashed) detected in the {\it Swift}/BAT survey with {\it Gaia} distances accurate better than
10$\%$. A hint of a bimodal distribution (solid line) in the IP sample 
could be present with four  low-luminosity systems overlapping the Polar sample.
 \label{fig:fig2}}
\end{figure}

The negligible absorption in the hard X-rays makes these
surveys unique for population studies. In particular the
{\it Swift}/BAT survey,
with a more uniform exposure over the sky was used to estimate the mCV
space densities but with large uncertainties \citep{pretorius13,reis13,pretorius14}
The {\it Gaia} DR2 release now offers the opportunity to
assess the true space densities. Using the shallow flux-limits
of the  70-month{\it Swift}/BAT sample of \cite{pretorius14} and the {\it Gaia} DR2
parallaxes, the IP space density results to be lower than previously estimated, 
with an upper limit of $\rm < 1.3\times10^{-7}\,pc^{-3}$ \citep{schwope18}. 
However, confirmation of this result needs larger samples from more sensitive 
surveys, such as that foreseen with the {\it eROSITA} satellite.  
Nevertheless, the recent
release  of the 105-month \citep{oh18} and the parallel 100-month \citep{cusumano14} {\it Swift}/BAT 
catalogues,  reaching flux levels down to
$\sim 7.2$($\rm \sim$5.4) and $\rm 8.4 \times 10^{-12}\rm erg\,cm^{-2}\,s^{-1}$
over 50 and 90$\%$ of the sky, respectively, are providing the opportunity to
enlarge the sample of hard X-ray detected CVs and in particular of mCVs. 
For the current sample of
hard X-ray mCVs, encompassing 50 confirmed IPs (see Table\,1) and 13 polars, 
we obtained distances from
the {\it Gaia} DR2 release  using a distance prior based on the Galaxy model described  
in \cite{bailer18} \footnote{http://gaia.ari.uni$-$heidelberg.de/tap.html}. 
The majority are accurate with relative uncertainties less than $10\%$, 
 restricting the sample to $\lesssim 1.8$\,kpc and $\lesssim$500\,pc
for 36  IPs and for the 13 polars, respectively.  
The derived distribution of 
hard X-ray luminosities in the {\it Swift}/BAT 14-195\,keV band is shown 
in Fig.\,\ref{fig:fig2} for this sample of mCVs. It peaks at
$\rm L_{hard} \sim 1.3\times 10^{33}\,erg\,s^{-1}$,  but also extends to low
luminosities ($\rm \lesssim  10^{32}\,erg\,s^{-1}$) where four systems are found, hinting
to a bimodality. The presence of a putative still-hidden 
faint ($\rm \sim 10^{31}\,erg\,s^{-1}$) population of IPs was already envisaged
by \cite{pretorius14} based on  the 70-month {\it Swift}/BAT catalogue. 
These low-luminosity IPs should accrete at  low rates and thus expected 
at short orbital periods (see also Sect.\,3). Three of the four low-luminosity 
IPs have indeed $\rm P_{\Omega} < 2\,h$.  The 13 confirmed hard X-ray polars 
detected so far 
\citep[see][]{bernardini14,gabdeev17,mukai17,bernardini17,bernardini19b}
are found at $\rm \lesssim 10^{33}\,erg\,s^{-1}$, overlapping the low-luminosity IPs. 
Among them, those few with determined magnetic field strengths 
(up to $\sim 40\times 10^6$\,G), not be expected to be strong in hard X-rays,
 challenge our knowledge of the emission properties in mCVs.
Whether the faint end of the mCV luminosity 
distribution is a mixture of the two classes or dominated by one of the two
groups needs to be assessed by classifying still unidentified 
faint sources at the survey flux limits.


\section{The increase of the mCV sample}\label{sec:results}

Both BAT and IBIS-ISGRI catalogues, still carry 
tentative new CV identifications, with many sources claimed as
magnetic, based on optical follow-ups 
\citep[e.g.][]{masetti12a,masetti13,parisi14,thorstensen13,thorstensen15}
and thus subject of revisions (Fig.\,\ref{fig:fig1}).
 Optical photometry may unveil coherent pulsations, although
these do not unambiguously identify the rotation period of the accreting WD 
(see Sect.\,3.1). Other cases occur when intermittent optical 
short-period variations are present that hamper a secure classification.
The detection of X-ray pulses and spectral 
characteristics instead efficiently diagnose the magnetically channeled
accretion flow onto the WD primary and thus allow 
firm confirmation of the magnetic status of candidates.


With a long-term programme using {\it XMM-Newton} we could confirm 
the magnetic status of 29 CVs, of which 26 resulted IP-type systems 
\citep[see][and references therein]{bernardini12,bernardini13,bernardini17,bernardini18,bernardini19}
and 3 of them were found  to be hard X-ray polars 
 \citep[][]{bernardini14,bernardini17,bernardini19b}.
We also disproved the mCV nature of 6 systems,  due to the lack of 
X-ray pulsations although for two of them X-ray spectra closely resemble
those of IPs \citep[][2019c,in prep]{bernardini13}. Noteworthy is the 
case of a bright hard X-ray low-mass X-ray binary, XSS\,J12270-4859, 
associated by us to a {\it Fermi}-LAT gamma-ray source, previously 
misidentified as an IP and later recognized
as one of the few intriguing transitional millisecond pulsar binaries 
\citep[see][]{demartino10,demartino13}. 

The current roster of confirmed IPs amounts to 69 systems with 
50 detected as hard X-ray sources (see Table\,\ref{tab:ips} for a complete
list of hard IPs as of September 2018)  
\footnote{Known and candidate systems can be found at 
https:://asd.gsfc.nasa.gov/Koji.Mukai/iphome/iphome.html}
 The increase in number of confirmed hard polars to 13, out of $\sim$130
\citep[][update RKcat7.24,2016]{ritter_kolb}, 
suggests they are not as rare as previously thought.

\subsection{Timing properties of identified mCVs}\label{subsec:timing}

The marking characteristics of mCVs is the presence of an X-ray periodic modulation at
the spin period of the WD. Therefore the synchronous polar systems are known 
to display marked (up to $\sim 100\%$) variability at the orbital ($\sim$hrs) period due 
to the self-occultation
of the accretion  spot behind the limb of the WD, 
giving rise to bright and faint
phases {bf \citep[see reviews by ][]{cropper90,mukai17}}\footnote{A handful 
of polars are found to be slightly ($\lesssim 2\%$) desynchronised, displaying also 
long-term variations at the sideband orbital frequency $\rm \omega - \Omega$  
\citep[see][and references therein]{ferrario15,rea17}}. The hard polars identified in 
our programme, Swift\,J2218.4+1925, Swift\,J0706.8+0325 and Swift\,J0658.0-1746,
  display similar modulation \citep[][]{bernardini14,bernardini17,bernardini19b} 
with the faint phases not reaching zero counts, indicating the presence of a 
secondary accreting pole. In two of them energy dependend orbital light curves 
are characterised by a narrow dip overimposed
on the bright phase, a common feature of polars, due to
photoelectric absorption when the  accretion stream
intercepts the line of sight \citep[e.g.][]{ramsay04a}. 
In Swift\,J2218.4+1925 this feature does not disappear at higher energies 
($>$ 2\,keV) and is similar to the rare case of EP\,Dra 
\citep[e.g.][]{ramsay04b},
indicating that there is a dense core obscuring the accretion region.
No soft X-ray additional component appears to be present in these new
polars (see also Sect.\,3.2). These systems are likely low-field polars 
($\sim 7-14\times 10^{6}$\,G) \citep[see details in][]{bernardini14,bernardini17}.

\noindent IPs instead display much faster periodic variability. Generally, 
the spin ($\rm P_{\omega}$) period dominates their X-ray power spectra indicating
accretion occurs via a disc \citep{wynnking92,norton96}. 
The presence of  sidebands and particularly a strong beat periodicity 
($\rm P_{\omega-\Omega}$) is the hallmark
of a disc-overflow accretion \citep{hellier95,norton96,norton97}. Only one
system, V2400\,Oph is known to date to display strong X-ray pulsations at the beat period, thus
representing an unique case among IPs where accretion onto the WD occurs withouth an
intervening disc \citep{buckley97,demartino04}. 

The new IPs also show a dominant spin periodicity
\citep{anzolin09,bernardini12,bernardini13,bernardini15,bernardini17,bernardini18}, with
a few also displaying weaker power at the harmonic of the spin frequency 
in their power spectra, indicating the presence of a secondary weakly emitting pole.  
Remarkable is the case of IGR\,J1817-2509, a pure two-pole accretor, displaying 
X-ray pulses only at 2$\omega$ \citep{bernardini12}. These systems are then 
disc-accretors, where the material from the disc attaches
to the magnetic field lines and is channeled onto the WD magnetic poles in an 
arc-shaped curtain \citep{rosen88}. In this configuration the maximum of the pulsation
is observed when the curtain points away from the observer and when the optical
depth of the accretion funnel is minimum \citep[][]{rosen88,nortonwatson89,norton99}. 
Most IPs have therefore the energy dependent spin pulses with amplitudes 
increasing at lower energies, indicative of photoelectric absorption along the pre-shock 
accretion flow. The absorbing material partially covers the X-ray emitting region
and it is complex, requiring a distribution of covering fraction as a function of column
density \citep[see details in][]{done_and_magdziarz98,mukai17}.
Few exceptions are those low
accretion rate IPs where absorption in the pre-shock flow is negligible and thus
the spin modulation is mainly due to changes in the visible portion of the emitting region
\citep[e.g.][]{allan98,demartino05}.
It has also to be noted that those few systems observed at energies $\geq$10\,keV 
at adequate S/N, such as with {\it NuSTAR}, have shown hard X-ray spin pulses 
that cannot be ascribed to absorption, implying non-negligible shock heights 
\citep{demartino01,mukai15}.

\noindent Noteworthy is the high incidence ($\sim 45\%$) of systems found to be 
spin-dominated
in the X-rays but dominated by the beat in the optical band \citep{bernardini12}, 
implying that the optical light is affected by reprocessing at fixed regions in the
binary frame. This also demonstrates that optical pulses 
are not reliable tracer of the WD rotation.
Many of the identified IPs  have been found to also display 
non-negligible variability at the beat ($\omega-\Omega$) frequency in the X-rays, 
which, in a few cases, can reach amplitudes as large as those of the spin 
modulation \citep[e.g.][]{bernardini17}. These systems
have an hybrid geometry where substantial portion of the accreting matter overflows the 
disc and directly impacts onto the WD magnetosphere. 

\noindent The long uninterrupted exposures with {\it XMM-Newton} 
 increased the 
number of IPs displaying substantial X-ray orbital variability. 
A few (8 so far) are eclipsing IPs,  giving the opportunity to study in details
the accretion geometry \citep[see][]{hellier14,esposito15,johnson17,bernardini17}.
Orbital modulations were found to be energy dependent from {\it ASCA}  
and  {\it RXTE} observations in 7 systems and interpreted as photoelectric absorption
due to material located at the disc rim \cite{parker05}. The number of IPs 
showing energy dependent orbital modulations
 has now increased to 13 systems \citep{parker05,bernardini12,bernardini17,bernardini18}. 
The amplitudes are found to range  from 3-4$\%$ up to $\sim 100\%$ as in the extreme
cases of FO\,Aqr, Swift\,J0927.7-6945 and IGR\,J14257-6117.  
Drawing similarities with low-mass X-ray binaries displaying orbital dips, IPs 
showing  large orbital modulations should be seen at moderately high ($\gtrsim 60^o$)
binary inclinations, allowing azimuthally extended absorbing material to intercept
the line of sight \citep{parker05,bernardini18}.
Changes of the amplitudes with epochs were already detected by \cite{parker05} 
and later confirmed by  \citep{bernardini18}, possibly linked to 
changes in the mass accretion rate. However, such possibility has not found confirmation
yet.

\noindent Interesting case is IGR\,J19552+0044 found to display two close long periods 
of 1.69\,h and 1.35\,h interpreted as the orbital and spin periods, respectively 
\citep{bernardini13}.
These have been recently improved with a long optical campaign 
resulting in $\rm P_{\Omega}$=1.393\,h and 
$\rm P_{\omega}$=1.355\,h  \citep{tovmassian17}. The spin-to-orbit period ratio is
remarkably high,  
$\rm P_{\omega}/P_{\Omega}$ = 0.97 making IGR\,J19552+0044 the IP with the lowest
degree of asynchronism, and joining ''Paloma" which has a spin-to-orbit period
ratio $\sim$0.83 \citep{joshi16}. 
These two systems are likely 
in their way to become polars and represent test cases for mCV evolution.

\subsection{Spectral properties of identified mCVs }\label{subsec:spectral}

The structure of the PSR has been the subject of  many  studies over the years. 
Detailed one-dimensional two-fluid hydrodynamic calculations coupled
with radiative transfer equations for cyclotron and Bremsstrahlung were 
performed for different regimes, including the so-called bombardment regime occurring
at very low local mass accretion rates and at high magnetic field strengths
\citep{woelk96,fischer01}. In the latter case a standing shock does not
develop and the WD atmosphere is heated from below by particle bombardment. 
 Many  X-ray spectral models of mCVs have been developed that account for
temperature and gravity gradients, for cyclotron cooling (in the polars), 
solving one or
two-fluid hydrodinamic equations and dipolar field geometry
\citep[see][]{wu94,cropper99,canalle05,saxton07,hayashi14}. Modifications in the 
models to  account for finite size of magnetosphere, especially in the low-field IPs, have
recently been performed to obtain more reliable mass estimates 
\citep{suleimanov16,suleimanov19}.

\noindent The mCV spectra are characterised a by multi-tem\-pe\-ra\-tu\-re optically thin 
plasma, signified by the presence
of the 6-7keV iron complex with H-like {\sc Fe\,XXVI} (6.9\,keV) and He-like 
{\sc Fe\,XXV}  (6.7\,keV) K$_{\alpha}$ lines  as well as  weaker K-shell features
of less heavier elements, plus a neutral Fe K$_{\alpha}$ fluorescent component. 
The modeling of the X-ray spectra  allows tracing temperatures 
from $\sim$0.2\,keV up to  $\sim$20-40\,keV in the PSR, 
although a continuous distribution is not always required  
\citep[e.g.][]{demartino08,anzolin09} indicating that indeed the emergent 
spectrum is highly sensitive to local pressure and temperature across the flow. 
Additionally a Compton reflection continuum component emerging at high energies 
and arising from  the WD surface was investigated
by \cite{suleimanov08} and found to be  unimportant for low WD masses and/or for low
mass accretion rates. 
Recently, stringent constraints on the presence of a Compton reflection hump in mCVs
have been found using {\it NuSTAR} observations of three bright
hard X-ray IPs \citep{mukai15}. This component is usually not required in the
spectral fits to the low S/N average BAT and/or IBIS-ISGRI spectra of most 
IP systems. 
Reflection either at the WD surface or in the pre-shock flow is 
anyway  confirmed by  
the above mentioned fluorescent Fe at 6.4\,keV \citep{ezukaishida99}, found
to be ubiquitous in the spectra of mCVs, with equivalent widths (EW) in the 
range $\sim$100-250\,eV. In a few systems, spin-phase 
resolved spectra show an increase in the EW at spin minimum, indicating an origin at the
WD surface \citep{bernardini12,bernardini17}.

\noindent Modeling of X-ray spectra of mCVs, especially the IPs, also requires 
complexities at low
energies, due to the presence of complex absorption from neutral material 
located in the pre-shock flow with column densities reaching values as high as 
$\rm \sim 10^{23}\,cm^{-2}$ \citep[see discussion in][]{mukai17}. 
Phase-resolved spectroscopy of these systems including the newly identified mCVs 
indeed  reveals that the 
partial covering fraction of the local absorber 
changes along the spin cycle  producing the observed energy 
dependence of spin pulses 
\citep[e.g.][]{bernardini12,bernardini17}. An additional absorption component is required
in those IPs displaying energy dependent orbital modulations 
\citep{bernardini12,bernardini18}.
While mCVs were not previously known to show ionized 
absorption features in their X-ray spectra, {\it XMM-Newton} and {\it Chandra} grating
spectra have demonstrated at least in three IPs the presence of an {\sc OVII} 
absorption edge 
\citep{mukai01,demartino08,bernardini12}. This indicates that in some IPs the 
pre-shock flow can be also substantially ionized.

\begin{figure}
\centerline{
\includegraphics[width=0.75\columnwidth,angle=-90]{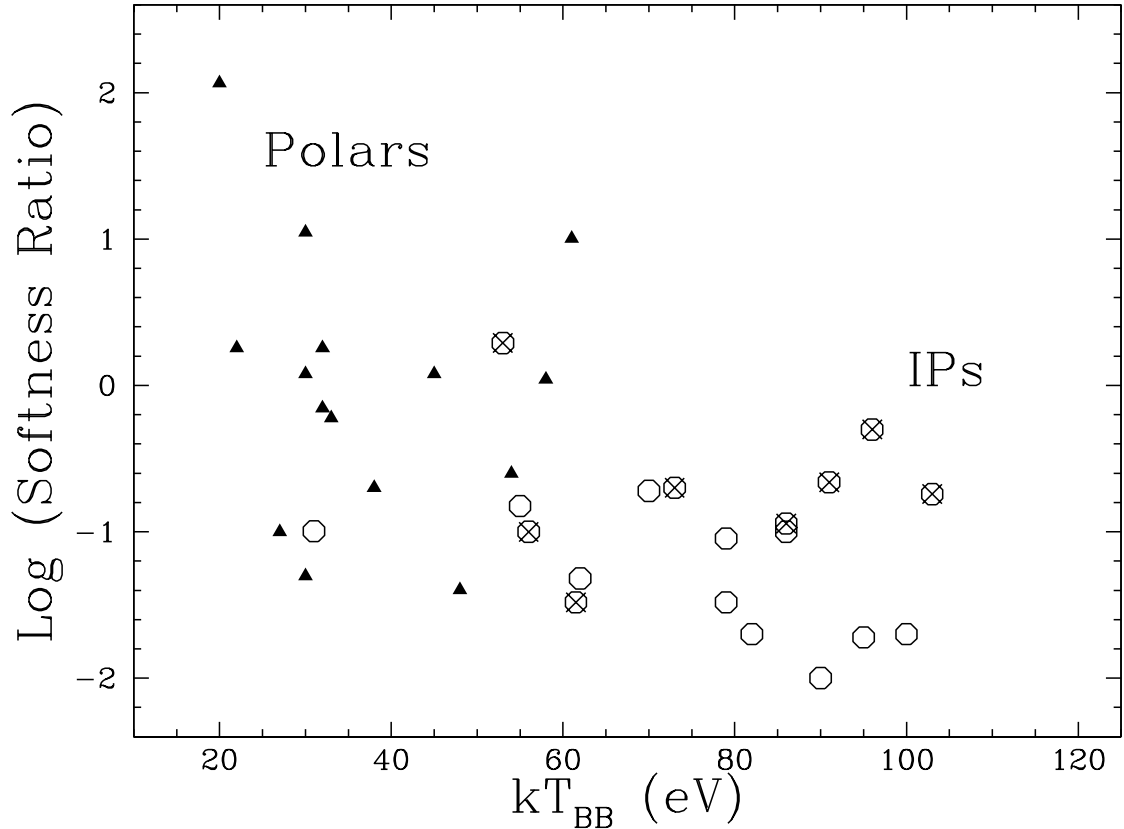}}
\caption{ The softness ratio of polars (filled triangles) and of soft IPs 
(open circles) versus blackbody temperature. The polarised IPs are also marked
with a cross. Adapted from \cite{bernardini17}.
 \label{fig:fig3}}
\end{figure}

\noindent Additionally, an optically thick soft ($\sim$20-60\,eV) component, arising from 
the heated polar cap, was believed to be the characterising spectral feature 
of polars \citep{beuermann99}, but not of IPs, except a handful of "soft" IP systems 
\citep{haberl_motch95,haberl02,demartino04}. 
{\it XMM-Newton} has remarkably shown 
an increasing number of polars without a distinct soft X-ray 
excess \citep{ramsay04c,ramsay04d,ramsay09,bernardini14,worpel16,bernardini17}, 
suggesting that if a reprocessed component exist, this has 
a low temperature shifting the emission towards the EUV/UV ranges. 
This also implies that polars cannot be characterised as soft X-ray emitting sources
anymore.
Thanks to the high sensitivity of {\it XMM-Newton} the
number of IPs showing a soft blackbody component has remarkably increased from the
three {\it ROSAT}-discovered systems to 19, representing
$\sim30\%$ of the whole IP class \citep[see details in][]{anzolin08,bernardini17}. 
Among them 8 are found to be polarised in the optical/nIR band \citep{ferrario15,potter18}, 
possibly suggesting that the detectability of the soft X-ray component could be
linked to the additional contribution of cyclotron radiation in the 
reprocessing at the WD surface ($\rm L_{BB} \sim  L_{X,hard} +  L_{cyc.}$).
The soft X-ray blackbody temperature are found to span a wider range
from $\sim$40\,eV up to $\sim$100\,eV with very different soft-to-hard X-ray luminosity
ratios (see Fig.\,\ref{fig:fig3}), but on average lower than those inferred
in the soft polars. The inferred fractional areas are much smaller than
those typically found in polars by two-three orders of magnitudes.
Although high blackbody temperatures arising from the WD surface would be 
locally super-Eddington, the possibility that the soft component originates 
instead in the coolest regions of the PSR above the WD surface is 
not supported by the spectral fits \citep{bernardini17}.
It could be also possible that, as demonstrated in the polar prototype AM\,Her,
there is a temperature gradient over a large area of the polar cap \citep{beuermann12},
with the inner core regions reaching such high temperatures. In the case of IPs, the
hotter regions are detectable against the high absorbing column rather than the cooler
and softer ones. Here we also note that due to the arc-shaped nature of the 
accretion spot in IPs, 
the structure of the heated area at the WD photosphere is likely to be different
than that in the polars. 


\section{The role of fundamental parameters}
We here discuss  some of the fundamental parameters 
obtained from the enlarged sample of IPs.

\subsection{The spin-orbit period plane of IPs}

\begin{figure}
\centerline{
\includegraphics[width=1.1\columnwidth]{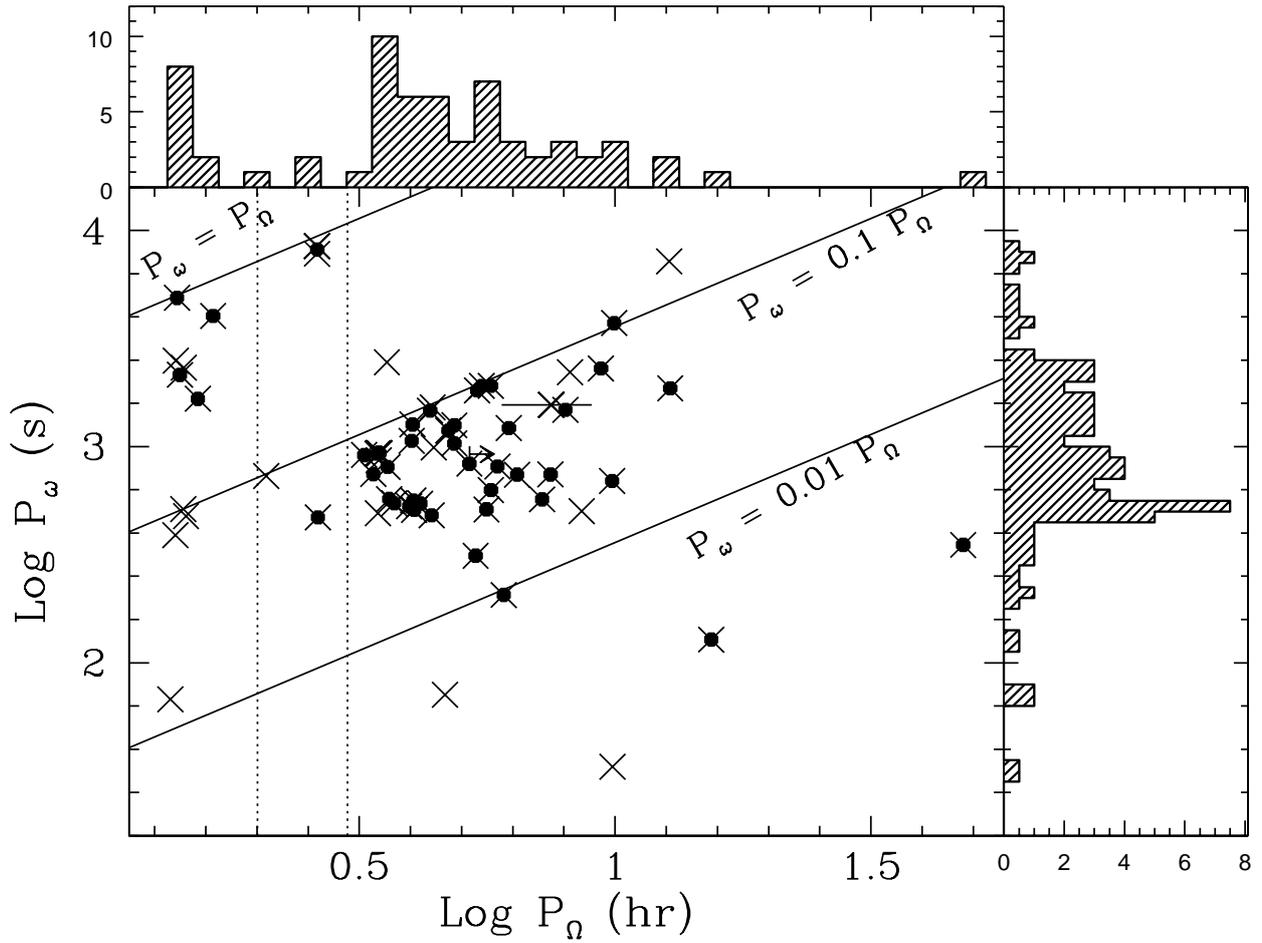}}
\caption{The spin-orbit period plane of confirmed IPs (crosses). 
Hard X-ray detected sources are also shown as filled circles. Solid lines
mark synchronism and two levels of asynchronism (0.1 and 0.01). Vertical
lines mark the orbital CV gap, where mass transfer is expected to stop at 
the upper bound and to be resumed at the lower bound. The spin (right
panel) and orbital (upper panel) period  distributions are reported 
(adapted from \cite{bernardini17}, including \cite{bernardini18,bernardini19} 
 \label{fig:fig4}}
\end{figure}

In Fig.\,\ref{fig:fig4} the spin-orbit period plane of the 69 confirmed 
IPs  with 
determined orbital periods is 
displayed together with the corresponding spin and orbital period distributions.
Previously known IPs were concentrated in a limited
range of the spin-orbit period plane, clustering just below $\rm P_{\omega}/P_{\Omega}
\sim 0.1$ with most systems found above the 2-3\,h orbital period gap, 
typically in the range 3-6\,h.  With the new identifications the 
plane has been substantially populated at long periods with 15 systems
at $\rm P_{\Omega}>$6\,h. Among them two are old novae, GK\,Per and 
SWIFT\,J1701.3-4304, recently identifed as 
Nova Sco AD\,1437 \citep[][]{shara17,bernardini17}. The new long period 
($\rm P_{\Omega}$=12.8\,h) IP, RX\,J2015.6+3711, still with an ambiguous identification
in  hard X-rays, is worth mentioning
for its extremely slow rotation ($\rm P_{\omega}$=2\,h) \citep{cotizelati16,halpern18},
only surpassed by "Paloma".   
These very long orbital period IPs represent
$\sim10\%$ of the whole CV population in this period range. 
Long period systems are believed to enter in the CV phase with nuclear 
evolved donors and may represent a significant portion of the present-day CV population 
\citep[][]{beuermann98,goliasch_nelson15}. Their spin-orbit period ratios, 
suggest they will reach synchronism while evolving to short orbital periods.
On the short period side, the number of IPs  below the 2-3\,h CV orbital period 
gap  has surprisingly increased to 10 members (Fig.\,\ref{fig:fig4}).
This is challenging since short period mCVs should have already 
reached synchronism if their magnetic moments are greater than 
$\rm \sim 5\times10^{33}\,G\,cm^3$ 
\citep[see][]{norton04,norton08}. The large spread in spin-to-orbit period
ratios of these short period IPs may suggest they belong to 
a different population of old, possibly, very low-field
systems that will never synchronise. 

\noindent The spin-orbit period ratios observed in IPs were investigated
in terms of their complex accretion geometry. \cite{norton04,norton08} showed
that if the WDs in IPs are spinning at equilibrium, 
systems with very small spin-orbit period ratios 
($\rm P_{\omega}/P_{\Omega}\lesssim 0.1$)  should accrete via a disc,  
while those ratios between 0.1 and 0.6 would be accreting via disc/stream, 
depending on the binary mass ratio, and reaching a ring-like configuration at 
$\rm P_{\omega}/P_{\Omega}\sim0.6$. Those IPs with higher spin-orbit period ratios 
should be far from equilibrium.


\begin{figure*}
\begin{center}
\hbox{
\includegraphics[angle=0,width=80mm,height=6.5cm]{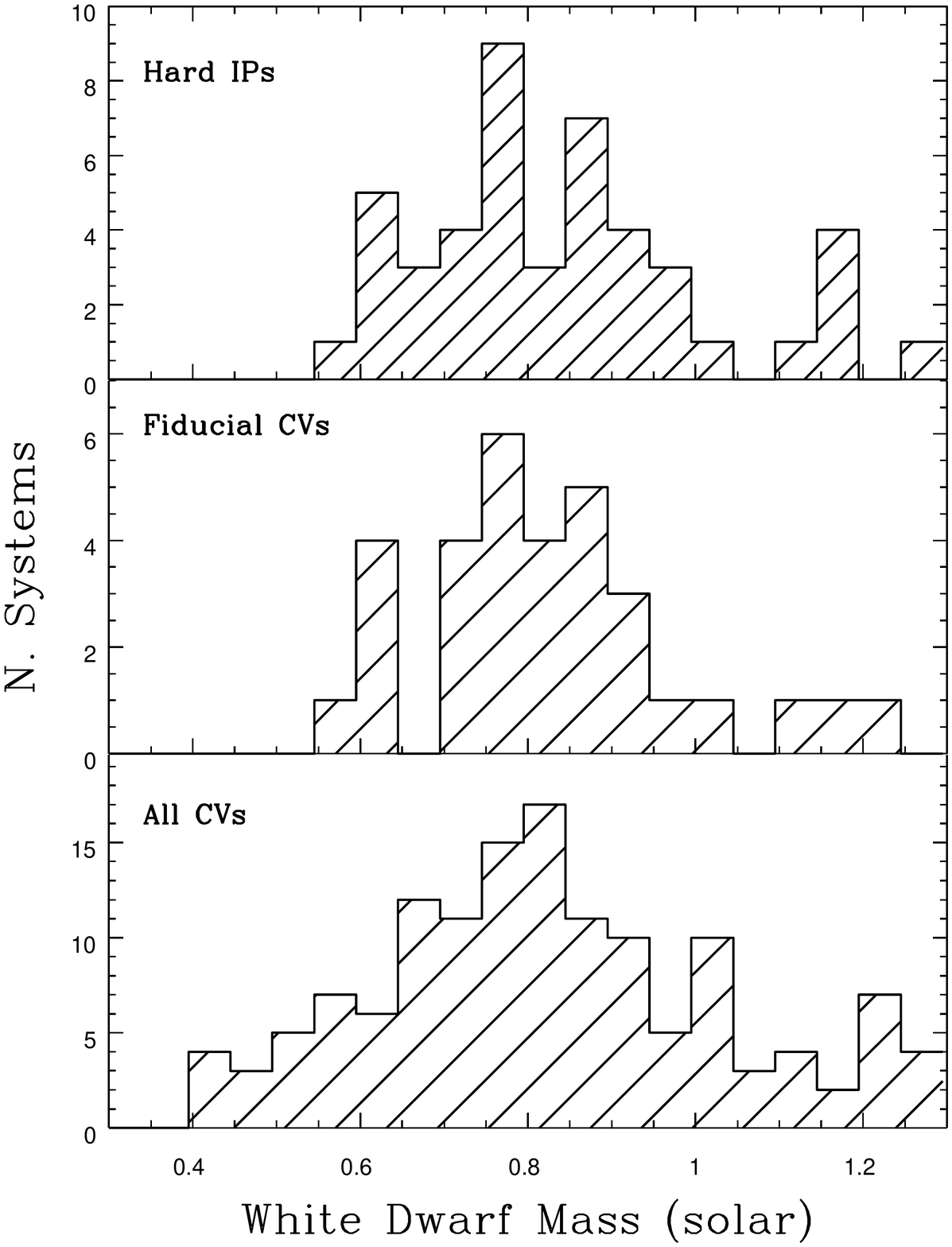}
\hspace{0.3cm}
\includegraphics[width=80mm,height=6.5cm]{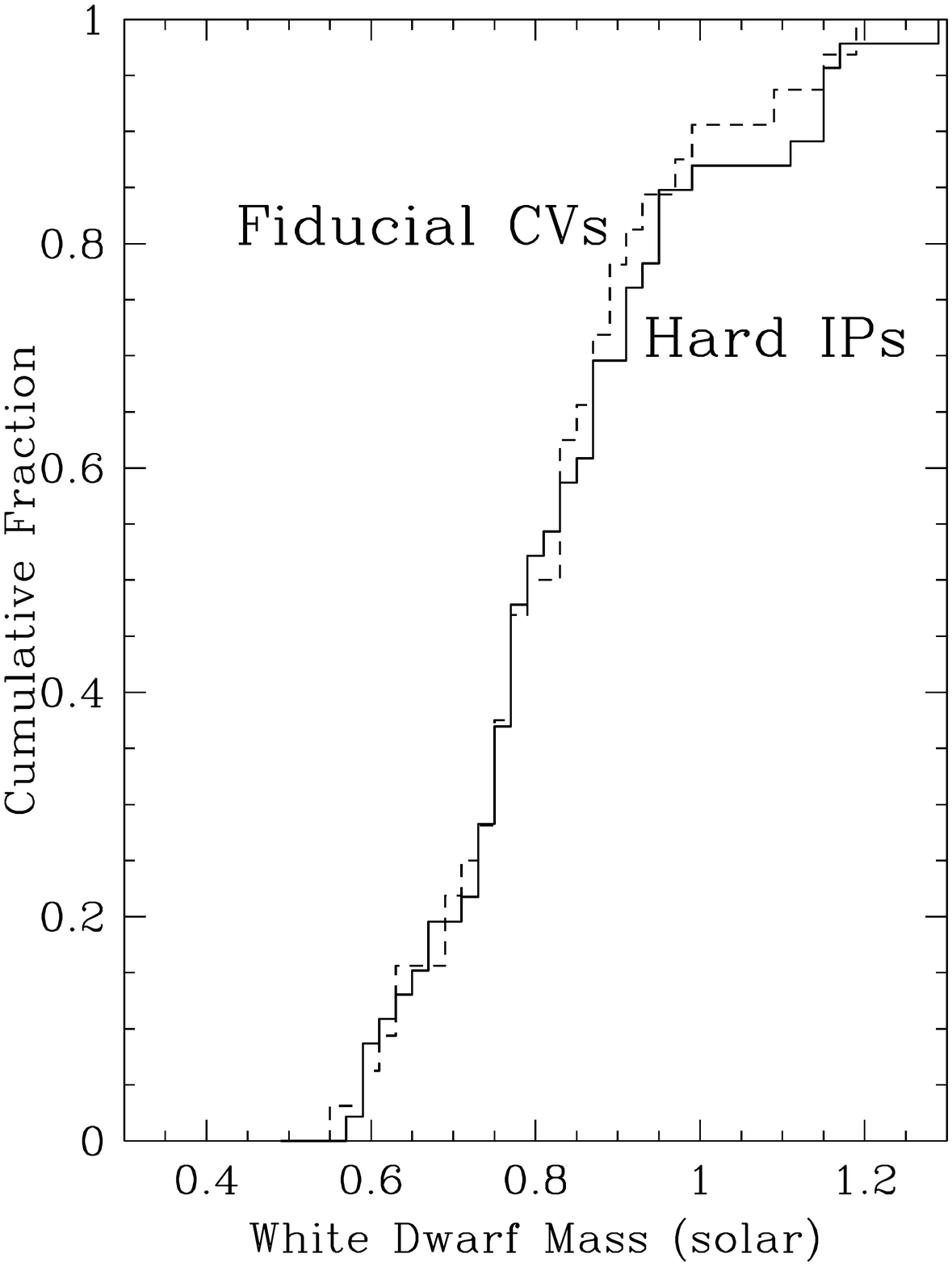}}
\end{center}
\caption{{\it Left panel:} The distributions of the WD masses of hard IPs (up) 
listed in Table\,1, of the fiducial CVs taken from \cite{zorotovic11} (middle) 
and those listed in the 
\cite[][update RKcat7.24,2016]{ritter_kolb} catalogue of CVs (bottom).
{\it Right panel:} The cumulative distributions of the mass of hard X-ray IPs 
(solid line) and the mass of the fiducial CVs (dashed line).
\label{fig:fig5}}
\end{figure*}


\subsection{The mass distribution of hard X-ray IPs}

The broad-band X-ray spectra of mCVs identified in our programme,  obtained
combining {\it XMM-Newton} and {\it Swift}/BAT or {\it INTEGRAL}/IBIS-ISGRI 
data,   increased the  
 sample of IP systems for which the WD mass has been 
estimated \citep{bernardini12,bernardini13,bernardini17,bernardini18,bernardini19}. 
These determinations are based on the PSR model by \cite{suleimanov05}. 
This model, and similar methods based on multi-temperature fit, have also 
been applied to the previously known bright hard IPs identified in the 
first 2.5\,yrs of {\it Swift}/BAT survey \citep{Brunschweiger09} and to a few of them
recently re-observed in the hard X-rays at much higher S/N with 
{\it NuSTAR} \citep{tomsick16,hailey16,shaw18,wada18}. The WD mass distribution of 
46 hard X-ray IPs is shown in the left upper panel of Fig.\,\ref{fig:fig5}, with a
mean value $\rm <M_{WD}>= 0.84\pm0.17\,M_{\odot}$. 
The modified PSR model accounting for the finite size of
the magnetosphere was also recently applied to a set of 35 IPs observed 
with {\it NuSTAR} and {\it Swift}/BAT by 
\cite{suleimanov19}, who find more accurate WD masses with an average value of 
0.79$\pm$0.16\,$\rm M_{\odot}$, but still consistent with previous results.
Massive WD primaries were also found 
by \cite{zorotovic11}, using a set of ''fiducial" CVs with reliable WD
masses (left middle panel of Fig\,\ref{fig:fig5}) with a mean value 
$\rm <M_{WD,fiducial}>= 0.82\pm0.15M_{\odot}$. Similar results are found using 
the WD masses listed in the \cite[][update RKcat7.24,2016]{ritter_kolb} 
catalogue of CVs (left bottom panel of Fig\,\ref{fig:fig5}).
We performed a Kolmo\-go\-rov-Smirnov (K-S) test between the masses of 
hard X-ray IPs and those of fiducial CVs, resulting in a 
probability of 99.3$\%$ 
that the distributions are from
the same parent population (right panel of Fig\,\ref{fig:fig5}). 
Hence, irrespective of being magnetic or not, WD primaries in CVs are 
more massive than single WDs ($\rm 0.6\,M_{\odot}$) and 
WDs in pre-CV binaries ($\rm 0.67\,M_{\odot}$).
 This could suggest that  WDs in CVs grow in mass during their 
evolution \citep[see][]{zorotovic11}, and that CVs may be favourable 
SN\,Ia progenitors. This result however strongly disagrees with 
the predictions of classical nova models \citep[e.g][]{prialnik95}.
Whether novae gain or loose mass is still highly controversial  
\citep[][]{yaron05,starrfield09} since the determination of the 
ejected mass is extremely challenging \cite[see][]{shara10,pagnotta15}. 

We also note that the mean mass of high field isolated magnetic WDs 
($\rm B\gtrsim10^6$\,G) is $\rm 0.78\pm0.05\,M_{\odot}$ very different from
the mean mass of single WDs \citep{ferrario15}.   
The fact that no magnetic WD is found in detached MS+WD binaries 
led \cite{tout08} and later 
\cite {wickramasinghe14} to propose the possibility that the strong 
differential rotation expected to occurr during the commone envelope 
(CE) phase may lead to the generation, by the dynamo mechanism, of a magnetic 
field that becomes frozen into the degenerate core of the pre-WD in MCVs making
these binaries hardly detectable. It is expected that at the end of the CE 
phase CV systems with smaller orbital separations  have WDs with stronger 
magnetic fields. Those systems that instead do merge are the   
progenitors of the single high field WDs.


\begin{figure*}
\begin{center}
\hbox{
\includegraphics[angle=0,width=80mm,height=6.5cm]{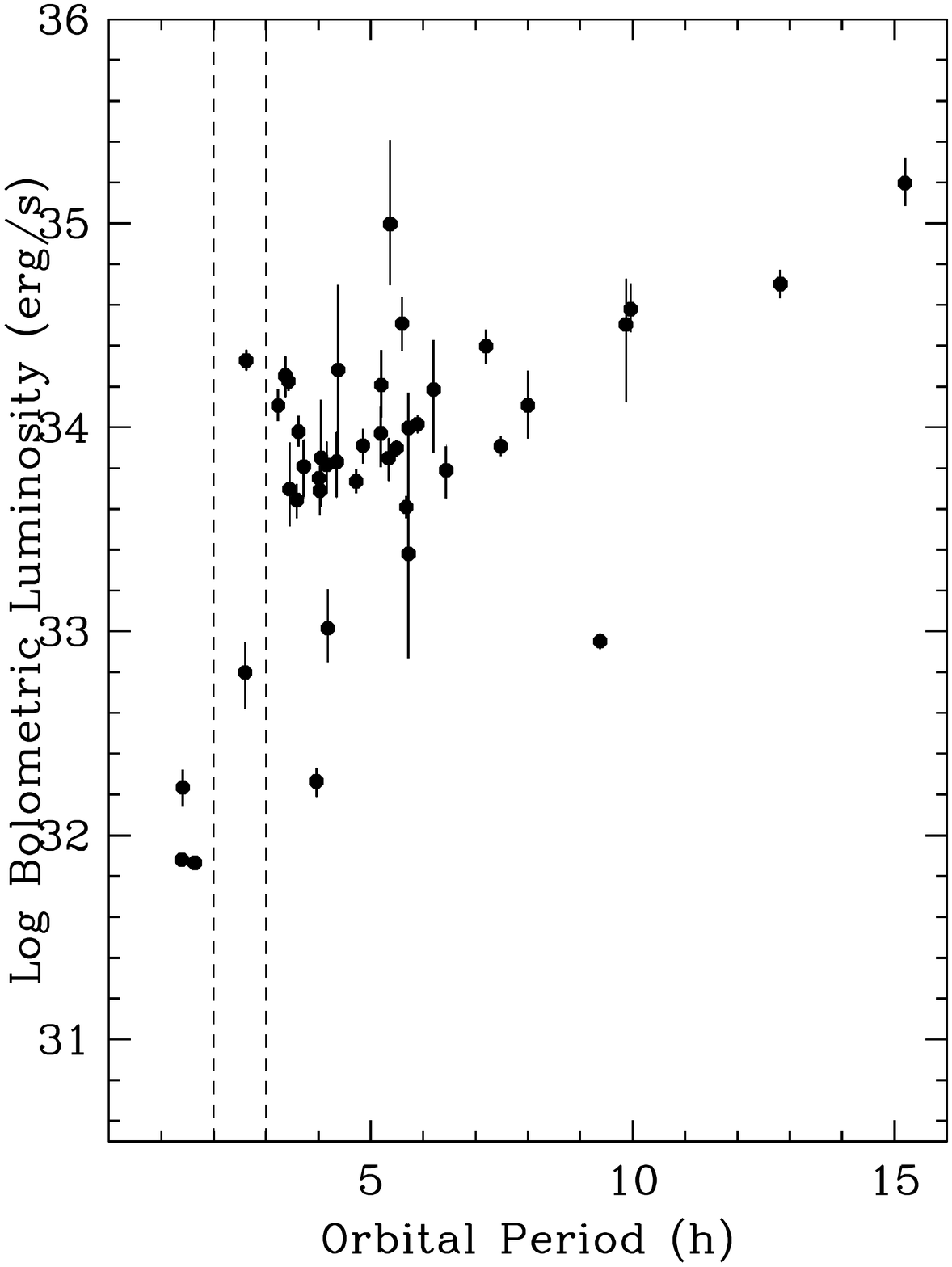}
\hspace{0.3cm}
\includegraphics[width=80mm,height=6.5cm]{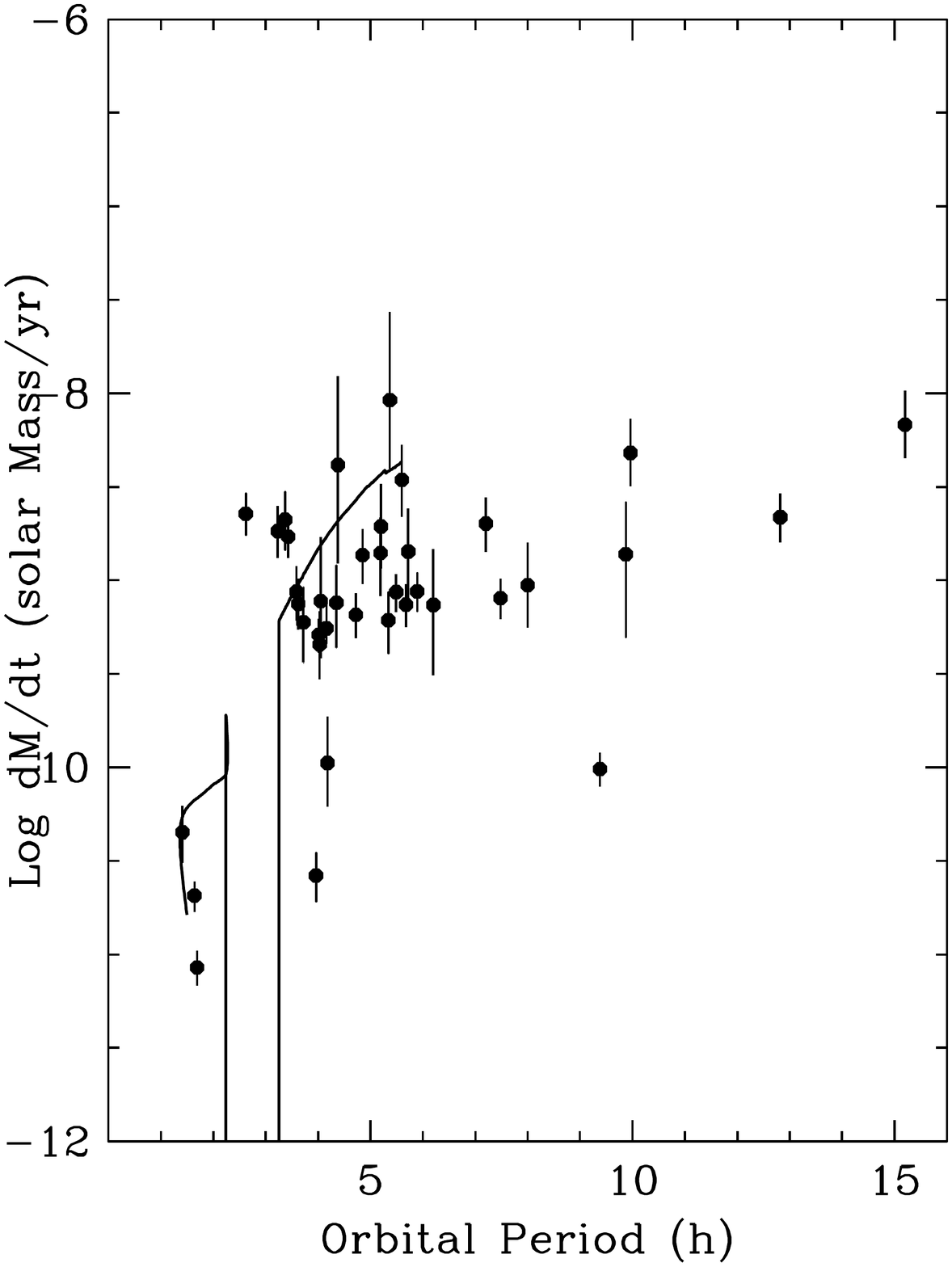}}
\end{center}
\caption{{\it Left panel:} The X-ray bolometric luminosities of 43 hard X-ray 
IPs with known orbital period, obtained using X-ray fluxes and distances 
reported in 
Table\,1. The dashed vertical lines mark the 2-3\,h period gap.
{\it Right panel:} The mass accretion rates versus $\rm P_{\Omega}$ 
as derived from the bolometric 
luminosities and adopting the WD masses reported in Table\,1 for 39 IPs. 
The solid line represent the revised CV evolutionary sequence by \cite{knigge11} for an assumed WD mass of 
0.75$\rm M_{\odot}$.
\label{fig:fig6}}
\end{figure*}


\subsection{The mass accretion rate }

The study of the broad-band spectra of IPs  allows a derivation of
the X-ray bolometric luminosity, also accounting for the reprocessed X-ray 
emission in the  soft IPs. The main
uncertainty in previous luminosity determinations  was  the distances,  
generally estimated with indirect methods \citep[e.g][]{bernardini12,bernardini17}. 
With the {\it Gaia} DR2 parallax release
it has been possible to obtain reliable distances. Most of the mCVs are
found to be at a larger distance than previously determined. We have used
a sample of 43 hard IPs with known orbital periods, distances and determined 
bolometric fluxes, reported in Table\,1. 
To evaluate the uncertainty in the luminosities we account for the errors in 
the derived  0.1-100\,keV X-ray fluxes and, when not available we account for
a $30\%$ uncertainty, and use the upper and lower bounds of 
distances derived with the method reported in \cite{bailer18}.
In Fig.\,\ref{fig:fig6} (left panel), we show
the luminosities for this sample of IPs as a function  of the 
orbital period up to 16\,h (only GK\,Per is out of the graph). 
The long period systems, as expected,  have larger luminosities by two 
orders of magnitude than the short period ones. 
This suggests that, if the X-rays trace the mass accretion rate, their
bolometric luminosities could be considered as proxies of the accretion 
luminosity. However,  the UV/optical light in IPs may carry
a non-negligible fraction of X-ray reprocessed flux \citep[see][]{mukai94}  
and thus the results have to be taken with  caution.
Assuming $\rm L_{X,Bol.} \sim L_{acc} = G\,M_{WD}\,\dot M\,R_{WD}^{-1}$ and using
the WD masses reported in Table\,1 and the M-R relation by \cite{nauenberg72},
we derive the mass accretion rates for a sample of 39 systems.
These rates versus $\rm P_{\Omega}$ are shown in the right panel of 
Fig.\,\ref{fig:fig6}, 
together with the  revised evolutionary sequence for an assumed WD  mass of  
0.75$\rm M_{\odot}$ obtained by \cite{knigge11}, adopting
a scaled version of angular momentum loss (AML) recipes for magnetic braking and
gravitational radiation. While for the polars the magnetic
coupling between the primary and the donor could be important in reducing
the efficiency of magnetic braking, thus lowering
the mass accretion rate
\citep{wickramasinghe_wu94,wickramasinghe00,ferrario15}, for the asynchronous
IPs this effect might not be important. The decrease in mass accretion rate
towards short orbital periods is broadly consistent with CV evolutionary theories
\citep{howell01,andronov03,knigge11}, although the long-period systems appear
to have lower rates than the predicted secular values. This was also noticed
in non-magnetic CVs using the effective WD temperature, $\rm T_{WD,eff}$, which is
a more robust $\rm \dot M$-proxy \citep{pala17}. 
Whether the accretion light or $\rm T_{WD,eff}$  are reliable tracers of 
donor mass transfer rates is  matter of  debate \citep{knigge11}.
The newly identified systems at very long periods,  
$\rm P_{\Omega} > $ 6\,h, should have donors that are nuclear evolved. Very long
orbital period CVs might represent a non-negligible fraction of 
present-day CV population still to be identified
\citep{goliasch_nelson15}.

Two long period systems, YY\,Dra and Swift\,J0746.3-1608 stick below the
bulk of IPs, indicating they have been caught in a low or intermediate luminosity
state. However, the majority of IPs are persistent
systems. Exceptions are EX\,Hya, HT\,Cam,
XY\,Ari, YY\,Dra and  GK\,Per that display occasional dwarf nova outbursts lasting
a few days except for GK\,Per whose outbursts last a few months. TV\,Col and 
V1223\,Sgr have instead shown hrs-long outbursts 
\citep[see][and references therein]{szkody02,hellier14}. 
These outbursts are due to disc instabilities.
In Fig.\,\ref{fig:fig6} none of them are reported during outburst. 

Two IPs, AO\,Psc and V1223\,Sgr have instead shown  
low accretion states in the past, with V1223\,Sgr remarkably 
dimming for about a decade  \citep{garnavich88}. Their orbital periods fall
in the 3-4\,h range where VY\,Scl stars, the novalike systems 
undergoing deep low states, are found 
\citep[see][]{rodriguez_gil07}. The orbital period of YY\,Dra (3.96\,h) falls
in the same range.  As for systems outside the VY\,Scl period range,
the longer period (4.85\,h)  IP, FO\,Aqr, was persistently 
observed at about the same level for several decades until 2016 when it underwent three 
low states in about 2\,yrs \citep{kennedym17,littlefield18}. \cite{littlefield19}
additionally find two low states occurring in the late 60-ties and mid 70-ties. 
The 
very long period (9.38\,h) system Swift\,J0746.3-1608 could not be identified
as an IP in an {\it XMM-Newton} observation in 2016 \citep{bernardini17} due to
its extreme faintness in the X-rays. 
It has been recently identified in 2018, when it recovered 
a strongly variable higher state after a possibly 6yrs-long faint X-ray level
\citep{bernardini19}.  Its position in the $\rm \dot M$-$\rm P_{\Omega}$ plane
suggests that it was still in a sub-luminous state in 2018.
The  low states are believed to be due to
a temporary reduction of the mass  transfer rate from the donor star
\citep{livio_pringle94}, causing the accretion disc to be dissipated once
the mass transfer drops below a critical threshold \citep{hameury_lasota17}.
In the well studied case of FO\,Aqr both the X-ray and
optical/UV modulations were found to vary in  amplitude over the years and 
also during the recent low states. This indicates that changes in the accretion 
mode are also likely driven by mass transfer rate
variations \citep{beardmore98,demartino99,kennedy17,littlefield19}.
State transitions, although rare in IPs, are an important
new aspect to understand  angular momentum loss governing the evolution of mCVs 
towards short orbital periods.

\begin{table*}
\addtolength{\tabcolsep}{-3.50pt}
\centering
\small
\caption {\leftline{ Parameters of confirmed hard X-ray IPs}}
\vskip 0.2cm
\begin{tabular*}{\textwidth}{l @{\extracolsep{\fill}} ccccccc}

\hline
 Name & P$_{\omega}$ & P$_{\Omega}$ & WD Mass  & $\rm F_{X,bol}^a$ & $\rm Distance^b$  & References\\
  &    (s)       &   (min) & ($\rm M_{\odot}$) &                    & (pc) &      \\ 
\noalign{\smallskip}
\hline
\noalign{\smallskip}
SWIFTJ0023.2+6142/V1033\,Cas & 563.5 & 242.0 & 0.91$^{+0.14}_{-0.16}$ & 1.84$\pm$0.2  & 1493$^{+137}_{-116}$ & 1,2,3,4\\
SWIFTJ0028.9+5917/V709\,Cas & 312.8 & 320.0 & 0.88$^{+0.05}_{-0.04}$ & 11.07$\pm$2.2 & 731$^{+12}_{-11}$ & 1,5,6,4\\
SWIFTJ0055.4+4612/V515\,And & 465.5 & 163.9 & 0.79$\pm$0.07 & 18.6$\pm$0.4   & 978$^{+47}_{-42}$ & 1,7,4\\
SWIFTJ0256.2+1925/XY\,Ari & 206.3 & 363.9 &  0.96$\pm$ 0.12 & $\cdots$ & $\cdots$ & 1,2 \\
SWIFTJ0331.1+4355/GK\,Per & 351.3 & 2875.4 & 0.87$\pm$0.08 & 5.5$^{+0.5}_{-0.9}$$\rm ^c$ & 437$^{+8}_{-10}$ & 1,8,4\\
SWIFTJ0457.1+4528         & 1218.7 & 371.3: & 1.12$\pm$0.06 & 3.2$\pm$0.2 & 2000$^{+570}_{-556}$ & 9,10,4\\  
SWIFTJ0502.4+2446/V1062Tau & 3704 &  598.9 & 0.72$\pm$0.17 & 13.90$\pm$0.48 & 1512$^{+209}_{-164}$ & 1,6,4\\
SWIFTJ0524.9+4246/Paloma & 7800: & 156 & $\cdots$ & 1.6$\pm$0.4  & 573$^{+37}_{-34}$ & 11,4\\
SWIFTJ0525.6+2416         & 226.3 & $\cdots$ & 1.01$\pm$0.06 & 5.0$\pm$0.4 & 1888$^{+362}_{-267}$ & 10,4\\
SWIFTJ0529.2-3247/TV\,Col & 1909.7 & 329.2 & 0.78$\pm$0.06 & 25.9$\pm$2.0 & 505$^{+5}_{-4}$ & 1,2,6,4\\
SWIFTJ0543.2-4104/TX\,Col & 1911 & 343.2 & 0.67$\pm$0.10 &  10.3$\pm$4.2 & 899$\pm$26 & 1,2,6,4\\
SWIFTJ0558.0+5352/V405\,Aur & 545.4 & 249.6 & 0.89$\pm$0.13 & 12.5$\pm$3.0$\rm ^d$ & 662$^{+14}_{-13}$ & 1,2,6,4\\
SWIFTJ0625.1+7336/MU\,Cam & 1187.2 & 283.1 & 0.74 $\pm$0.13 & 5.0$\pm$0.4$\rm ^d$ & 954$^{+26}_{-25}$ & 1,2,12,4\\
SWIFTJ0636.6+3536/V647\,Aur & 932.9 & 207.9 & 0.74$\pm$0.06 & 3.1$\pm$0.3 & 2073$^{+339}_{-260}$ & 1,7,4\\
SWIFTJ0704.4+2625/V418\,Gem & 480.7 & 262.8 & 0.5$\pm$0.2 & 2.46$\pm$1.0 & 2550$^{+927}_{-602}$ & 1,13,4\\ 
SWIFTJ0731.5+0957/BG\,CMi & 913.5 & 194.1 & 0.67$\pm$0.19 & 11.5$\pm$0.8 & 966$^{+56}_{-50}$ & 1,2,6,4\\   
SWIFTJ0732.5-1331/V667\,Pup & 512.4 & 336.2 & 0.79$\pm$0.11 & 7.9$\pm$1.0 & 1848$^{+178}_{-150}$ & 1,2,4\\ 
SWIFTJ0746.3-1608           & 2311: & 562.8 & 0.78$\pm$0.13 & 1.84$\pm$0.06 & 638$\pm$12 & 14 \\
SWIFTJ0750.9+1439/PQ\,Gem  & 833.4 & 311.6 & 0.65$\pm$0.09 & 13.9$\pm$3.9$\rm ^d$ & 750$^{+21}_{-20}$ & 1,2,6,4\\
2PBCJ0801.2-4625          & 1306.3 & $\cdots$ & 1.18$\pm$0.10 & 4.8$\pm$0.3 & 1315$^{+48}_{-49}$ & 15,4\\
SWIFTJ0838.0+4839/EI\,UMa & 741.6 & 386.1  & 9.1$^{+0.017}_{-0.07}$$\rm ^e$ & 4.3$\pm0.9\rm ^e$ & 1095$^{+47}_{-43}$ & 1,4\\
IGRJ08390-4833/SWIFTJ0838.8-4832   & 1480.8 & 480:  & 0.95$\pm$0.08 & 2.52$\pm$0.3 & 2064$^{+311}_{-240}$ & 1,7,4\\
SWIFTJ0927.7-6945         &  1033.5 & 291.0 & 0.58$\pm$0.10 & 2.7$\pm$0.16 & 1123$^{+39}_{-38}$ & 15,16,4\\ 
SWIFTJ0958.0-4208         & 296.2 & $\cdots$ & 0.74$\pm$0.11 & 2.2$\pm$0.13 & 1586$^{+161}_{-134}$ & 15,4\\
SWIFTJ1142.7+7149/YY\,Dra & 529.3 & 238.1 & 0.75$\pm$0.02 & 3.9$\pm$0.6 & 198$\pm$1 & 1,17,6,4\\
SWIFTJ1238.1-3842/V1025\,Cen & 2146.6 & 84.6 & 0.60$^{+0.06}_{-0.03}$ & 3.4$\pm$0.6 & 192$^{+5}_{-4}$ & 1,18,6, 4\\ 
SWIFTJ1252.3-2916/EX\,Hya & 4021.6 & 98.3 & 0.78$\pm$0.03 & 13.8$\pm$1.0 &  56.85$^{+0.05}_{-0.13}$ & 1,19,20,4\\
IGRJ14091-6108/Swift J1408.2−6113 & 576.3 & $\cdots$ & 1.2-1.3 & 1.1 & 2808$^{+1157}_{-673}$ & 21,4\\
IGRJ14257−6117/4PBCJ1425.1-6118 & 509.5 & 243 & 0.6$\pm$0.2 & 2.19$\pm$0.22 & 1645$^{+533}_{-329}$ & 6,4\\
IGRJ1509-6649/SWIFTJ1509.4-6649 & 809.4 & 353.4 & 0.89$\pm$0.08 & 6.8$\pm$0.2 & 1127$^{+37}_{-34}$ & 1,7,4\\
SWIFTJ1548.0-4529/NY\,Lup & 693.0 & 591.8 & 1.16$^{+0.04}_{-0.02}$ & 17.7$\pm$9.7 & 1228$^{+44}_{-40}$ & 1,5,6,4 \\
IGRJ16500-3307/SWIFTJ1649.9-3307 & 571.9 & 217.0 &  0.92$\pm$0.06 & 6.1$\pm$0.20 & 1140$^{+88}_{-77}$ & 1,7,4\\
IGR\,J16547-1916/SWIFTJ1654.7-1917 & 546.7 & 222.9 & 0.85$\pm$0.15 & 4.7$^{+0.3}_{-1.4}$ & 1066$^{+60}_{-55}$ & 1,22,4\\
SWIFTJ1701.3-4304/Nova Sco AD\,1437 & 1858.7 & 769.0 & 1.16$\pm$0.12 & 41.0$\pm$2.5 &  1014$^{+54}_{-48}$ & 15,4\\
SWIFTJ1712.7-241/V2400\,Oph & 927.7 & 205.8 & 0.81$\pm$0.10 & 28.6$\pm$1.8 & 701$^{+17}_{-16}$ & 1,2,6,4\\
IGRJ1719-4100/SWIFTJ1719.6-4102 & 1053.7 & 240.3 & 0.86$\pm$0.06 & 11.4$\pm$0.2 & 643$\pm$17 & 1,7,4\\
SWIFTJ1730.4-0558/V2731\,Oph  & 128.0 & 925.3 & 1.16$\pm$0.05 & 28.1$\pm$0.5 & 2165$^{+316}_{-245}$ & 1,23,24,4\\
AXJ1740.2-2903     & 628.6 & 343.3 & $\cdots$ & $\sim$1.1 & 1351$^{+1532}_{-474}$ & 1,25,4\\
IGRJ18173-2509/SWIFTJ1817.4-2510 & 1663.4 & 91.9: & 0.96$\pm$0.05 & 7.40$\pm$0.20 & $\cdots$ & 1, 4\\
IGRJ18308-1232/SWIFTJ1830.8-1253 & 1820.0 & 322.4 & 0.85$\pm$0.06 & 19.3$\pm$4.0 & 2074$^{+1232}_{-591}$ & 1,7,4\\
SWIFTJ1832.5-0863/AXJ1832.3-0840 & 1552.3 & $\cdots$ & $\cdots$ & $\sim$ 1.54  & 1051$^{+2421}_{-482}$    & 1,26,4\\
SWIFTJ1855.0-31/V1223\,Sgr &  745.5 & 201.9 & 0.75$\pm$0.02 & 46.0$\pm$7.6 & 571$^{+16}_{-15}$ & 1,5,6,4\\
IGRJ19267+1325 & 935.1 & 206.9 & $\cdots$ & $\sim$2 & 1443$^{+439}_{-276}$ & 1,27,4\\
IGRJ19552+0044/SWIFTJ1955.2+0077 & 4877.4 & 83.6 & 0.77$^{+0.02}_{-0.03}$ & 2.20$\pm$0.10 & 170$^{+3}_{-2}$ & 28,29,4\\
SWIFTJ1958.3+3233/V2306\,Cyg  &  1466.7 & 261.0 & 0.77$\pm$0.16 & 4.3$\rm \pm0.9^e$ & 1311$^{+61}_{-55}$ & 1,2,4\\
SWIFTJ2113.5+5422  & 1265.6 & 241.2 & 0.81$^{+0.16}_{-0.10}$  & 2.9$\pm$0.17 & 547$^{+116}_{-82}$ & 15,4\\
SWIFTJ2123.5+4217/V2069Cyg    & 743.1 & 448.8 & 0.82$\pm$0.08 & 5.20$\pm$0.20 & 1140$^{+43}_{-40}$ & 1,7,4\\    
IGRJ21335+5105/SWIFTJ2133.6+5105 & 570.8 & 431.6 & 0.93$\pm$0.04 & 11.9$\pm$1.5 & 1325$^{+48}_{-45}$ & 1,3,4\\
SWIFTJ2217.5-0812/FO\,Aqr   & 1254.5 & 290.9 & 0.61$\pm$0.05 & 25.38$\pm$3.64 & 518$^{+14}_{-13}$ & 1,2,6,4\\
SWIFTJ2255.4-0309/AO\,Psc   & 805.2 & 215.5 & 0.55$\pm$0.06 & 15.43$\pm$2.33 & 488$^{+11}_{-10}$ & 1,2,6,4\\


\noalign{\smallskip}

\hline
\end{tabular*}

\begin{flushleft}
$\rm ^a$:  Fluxes in units of $\rm 10^{-11} erg\,cm^{-2}\,s^{-1}$.
$\rm ^b$: Distances obtained from {\it Gaia} parallaxes and weak distance prior using 
Galactic model described in
\citep{bailer18}. Uncertainties encompass lower and 
upper bounds of highest posterior probability. Unreliable distances are not reported. 
$\rm ^c$: Quiescent flux. 
$\rm ^d$: Average flux between two observations.
$\rm ^e$: Obtained from combined {\it Swift}/XRT and BAT archival 
spectra retrieved from the products generator at the 
UK {\it Swift} Science Data Centre \citep{evans09} 
and from the NASA/GSFC {\it Swift}-BAT 105-month Web site 
https://swift.gsfc.nasa.gov/results/bs105mon/, respectively. \\

 


References: (1)$\rm P_{\omega},P_{\Omega}$ from \citet{ferrario15}; (2) \citet{Brunschweiger09}; 
(3) \citet{anzolin09}; (4) This work, (5) \citet{shaw18}; (6)\citet{bernardini18}; 
(7) \citet{bernardini12}; (8) \citet{wada18}; (9) \citet{thorstensen13}; (10) \citet{bernardini15};
(11) \citet{joshi16}; (12) \citet{staude08}; (13) \citet{anzolin08}; (14)\citet{bernardini19}; 
(15) \citet{bernardini17}; (16) \citet{halpern18}; (17) \citet{suleimanov05}; (18) \citet{ramsay00}; (19) \citet{echevarria16}; 
(20) \citet{luna18}; (21) \citet{tomsick16}; (22) \citet{lutovinov10}, (23) \citet{hailey16};
(24) \citet{demartino08}; (25) \citet{masetti12a}, (26) \citet{masetti13}; (27) \citet{masetti09};
(28) \citet{tovmassian17}; (29) \citet{bernardini13}; 

\end{flushleft}
\label{tab:ips}
\end{table*}


\section{Discussion and Conclusions}\label{sec:discussion}

We have presented the main results of an ongoing identification X-ray programme
of new mCVs discovered in the {\it INTEGRAL}/IBIS-ISGRI and 
{\it Swift}/BAT surveys,  that almost doubles  the current roster of 
IP-type systems and adds three new systems to the small group of 
hard X-ray polars.

\noindent Whether IPs are easily identified in these hard X-ray surveys because of 
massive WD primaries cannot be confirmed from their mass distribution, which 
results to be similar to that of other CVs. As demonstrated by
\cite{fischer01}, the PSR flow is one-fluid plasma in presence of 
moderate magnetic field strengths ($\rm B\lesssim30\times 10^{6}\,G$) 
and high flow rates ($\dot m \rm \gtrsim  1-5\,g\,cm^{-2}\,s{-1}$),  
 where radiative losses are mainly via Bremsstrahlung rather than cyclotron. 
It is then concievable that not only the IPs but also those polars 
with  high local mass accretion rates can be  detected in the harder X-ray
bands. 
While nowdays the multi-temperature structure of the PSR in mCVs can be
efficiently
 diagnosed in the brighter systems through grating spectra, 
the foreseen ESA large mission 
{\it Athena} \citep{nandra13} will  routinely perform such studies
 for a large number of fainter systems as well as  precisely 
measure WD masses via gravitational redshifts of Fe line.
The increasing number of polars without a soft X-ray excess and 
of  IPs displaying a soft optically thick  component indicates that 
the previous separation between the two
subclasses based on this spectral characteristics is  no longer valid.

\noindent The new idenfications have allowed to enlarge the range of orbital periods
of IPs,  with 10 systems below the gap and 15 above the the poorly explored
range $\rm P_{\Omega}\gtrsim$ 6\,h. A wide range of spin-orbit period ratios is 
found, with most short-period IPs below the 2-3\,h gap 
possessing a weak degree of asynchronism.
These IPs may represent a faint population of very weakly magnetised systems
that will likely never synchronise.
The advent of sensitive near-future survey experiments as 
{\it eROSITA} \citep{merloni12} or planned
such as {\it eXTP} \citep{intzand19} and {\it Theseus} \citep{amati18} 
will have a crucial role in unveling the 
true population of mCVs and in monitoring their still unexplored 
long-term X-ray behaviour.

\section*{Acknowledgments}

This work has been presented at the 42$^{nd}$ COSPAR Assembly in 2018 in Pasadena, 
USA, at  the Session entitled ''Nova Eruptions, Cataclysmic Variables and related systems: 
Observational vs. theoretical challenges in the 2020 era”. 
This work is  based on data obtained with 
{\it XMM-Newton} and {\it INTEGRAL}, ESA science 
missions  with instruments and contributions directly funded by ESA Member States, 
with {\it Swift}, a NASA science missions with Italian participation, with 
{\it NuSTAR}, a NASA science mission and with 
{\it Gaia}, an ESA mission, whose data are processed by the Data 
Processing and Analysis Consortium (DPAC).
DdM acknowledges financial support from INAF-ASI agreement I/037/12/0,
ASI-INAF contract n.2017-14-H.0 and INAF-PRIN
SKA/CTA Presidential Decree 70/2016. FB is
founded by the European Union’s Horizon 2020
research and innovation programme under the
Marie Sklo\-dow\-ska-Curie grant agreement n. 664931.
NM acknowledges financial support from ASI-INAF contract n.2017-14-H.0

\bibliography{biblio_last}


\end{document}